\documentclass[12pt, draftclsnofoot, onecolumn]{IEEEtran}
\usepackage{cite}
\usepackage{graphicx}
\usepackage[cmex10]{amsmath}
\usepackage{color,soul}
\begin{document}
	
	\title{Spatial-Mode Diversity and Multiplexing for FSO Communication with Direct Detection}

\author{Shenjie Huang, \IEEEmembership{Student Member, IEEE}, Gilda Raoof Mehrpoor, and Majid Safari, \IEEEmembership{Member, IEEE}}

\maketitle
	\maketitle
	\begin{abstract}
		This work investigates spatial-mode multiplexing (SMM) for practical free-space optical communication (FSO) systems using direct detection. Unlike several works in the literature where mutually incoherent channels are assumed, we consider mutually coherent channels that accurately describe SMM FSO systems employing a single laser source at the transmitter with a narrow linewidth. We develop an analytical model for such mutually coherent channels and derive expressions for aggregate achievable rate (AAR). Through numerical simulations, it was shown that there exist optimal transmit mode sets which result in the maximal asymptotic AAR at high transmitted power. Moreover, in order to resolve the reliability issues of such SMM FSO systems in the presence of turbulence, a so-called mode diversity scheme is proposed that can be easily implemented along with SMM FSO systems. It is demonstrated that mode diversity can significantly improve the outage probability and the $\epsilon$-outage achievable rate performance of the multiplexed channels in SMM FSO systems degraded by turbulence.    
		
	\end{abstract}
	
	\section{Introduction}\label{intro}
	Spatial-mode multiplexing (SMM) in free space optical communications (FSO) is the counterpart of the mode-division multiplexing (MDM) in fibre optics that has recently attracted more attention \cite{willner,wang}. Due to the orthogonality among beams with different spatial modes, they are proposed to be employed in communication systems to transmit multiple data streams simultaneously \cite{Anguita:08}. Similar to the traditional multiple-input-multiple-output (MIMO), SMM has the potential of achieving high degrees of freedom (DOFs) for communication \cite{zhao}. Recently, a number of spatial mode sets have been applied in FSO systems such as Laguerre-Gaussian (LG) beams \cite{Gibson:04} and Hermite-Gaussian (HG) beams \cite{zhao}. In particular, numerous works have been focused on orbital angular momentum (OAM) modes mainly because of the smaller space-bandwidth product and simpler generation and (de)multiplexing techniques, despite being only a subset of the complete LG basis \cite{zhao,willner}. 
	
Although theoretically SMM can boost the aggregate capacity, the performance of SMM FSO systems is impaired by the atmospheric turbulence \cite{Anguita:08}. After propagation through the atmosphere, the orthogonality can not be preserved and the reliability of communication might be significantly degraded \cite{Paterson}. In the long-haul fibre-based MDM communication systems, coherent detection and MIMO digital signal processing (MIMO-DSP) are usually employed to compensate crosstalk introduced by mode coupling \cite{Gasulla:15}. Many works on SMM FSO systems also use the same receiver scheme to mitigate the crosstalk caused by turbulence \cite{Huang:14,huang2015mode}. However, coherent detection is quite expensive and is not compatible with the requirement of low cost in practical FSO links \cite{survey}. Moreover, with large number of employed spatial modes the complexity of MIMO receivers is also an issue even in fibre optic systems, which leads to the partial MIMO or MIMO-free MDM systems \cite{Igarashi:15}. On the other hand, the application of adaptive optics on SMM FSO systems are also investigated \cite{Ren:14,shengmei}, however, this technique also significantly increases the link costs especially when large transceiver apertures are employed.
	
	Considering that the receivers with intensity modulation direct detection (IM/DD) are widely employed in practical terrestrial FSO links due to their simplicity, stability and low cost \cite{survey}, in this work we will focus on IM/DD SMM FSO systems.  In literature, mutually incoherent channels are usually assumed in such multiplexing systems to ensure the incoherent power addition between the intended signal and interference from other channels \cite{Anguita:08,Yadin:06}. With this assumption, the channel can be described as a linear MIMO channel with a positive-valued channel matrix and hence traditional MIMO-DSP can be applied to mitigate the crosstalk \cite{Nazarathy:08}. Two ways to realize this incoherent power superposition include generation of different transmitted spatial modes by distinct lasers with frequency differences larger than the receiver electrical bandwidth \cite{shane,Yadin,Legg} and using lasers with a linewidth much larger than the receiver electrical bandwidth \cite{Monroy}. In such cases, the interferometric noise (or beat noise) of the received optical power caused by the square-law photodetector characteristics can be averaged out and the system shows linear behaviour in the received optical power \cite{Monroy}. However, in both cases, the additional spatial DOFs of SMM are achieved in the expense of consuming more spectral DOFs than needed, which could be exploited through wavelength division multiplexing. Therefore, they do not correspond to an efficient design of SMM systems that aim to boost the data rate of FSO communication. 

	In order to simplify the transmitter design and preserve the spectral DOFs, a single laser source with narrow linewidth can be employed in MDM or SMM systems to generate the transmitted spatial modes. Since all multiplexed channels are originated from the same source, they are mutually coherent which results in the coherent superposition between the intended optical signal and the crosstalk at the receiver \cite{Yadin}. Due to the quadratic nature of the photodetectors, the channel description is now non-linear and traditional MIMO-DSP cannot be employed. Mutually coherent channels have been investigated especially in MDM systems with multi-mode fibres (MMFs) and some techniques such as zero-forcing beamforming \cite{Nazarathy:08,Bikhazi} and optical MIMO equalizer \cite{Arik:16,ArikarXiv} have been proposed to suppress the effect of crosstalk. 

	In this paper, we aim to investigate the performance of SMM FSO systems with mutually coherent channels impaired by both shot noise and thermal noise. Although mutually coherent channels have been studied in multi-mode fibers  \cite{Arik:16,Yadin} and near-field FSO multiplexing systems \cite{majidNF}, to the best of authors’ knowledge, IM/DD SMM FSO systems with such channels have not been investigated before. Moreover, by describing the detected signal based on a doubly stochastic Poisson model, we derive a novel expression for the aggregate achievable rate. In addition, in order to enhance the reliability of SMM FSO systems cost-effectively, a mode diversity scheme is proposed and studied. 
	
	The rest of the paper is organized as follows: In Section \ref{channel}, we introduce channel model for the investigated multiplexing systems. In Section \ref{perf}, we derive the average aggregate achievable rate (AAR) for such systems and discuss the optimal transmitted mode set which leads to the maximal asymptotic AAR at high transmitted power. In Section \ref{MD}, mode diversity is proposed for reliability improvement and the corresponding outage performance is presented. Finally, we conclude this paper in Section \ref{conc}. 

	\section{Channel Model}\label{channel}
	
\begin{figure}[!t]
			\centering
			\includegraphics[width=0.8\textwidth]{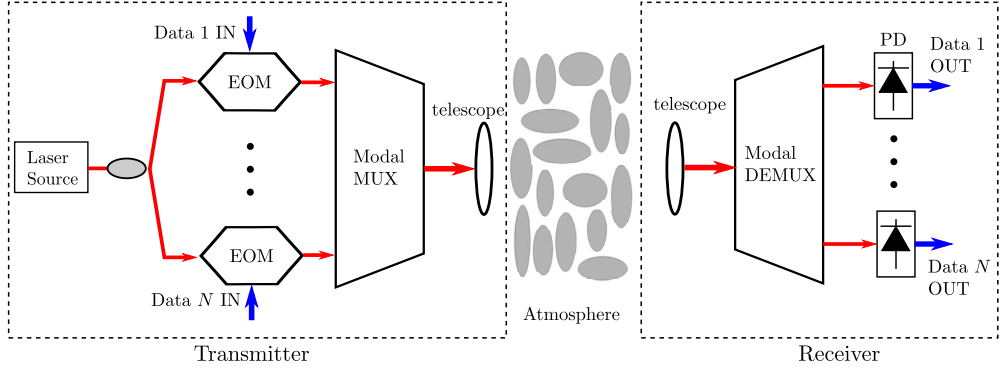}
			\caption{FSO SMM system with mutually coherent channels. EOM: electro-optic modulator; PD: photodetector.  }\label{scheme}
		\end{figure}

	\figurename\textrm{ }\ref{scheme} shows the schematic of the FSO SMM system with mutually coherent channels. At the transmitter, a single laser source with a narrow linewidth is employed. The electro-optic modulators (EOMs) are used to modulate $N$ input data streams onto the split beams. The modulated beams are converted into $N$ orthogonal spatial modes and the multiplexed beam is then transmitted through the transmitter telescope. At the receiver, the received optical beam is firstly demultiplexed to separate different spatial modes concerned and these modes are all converted back to the fundamental Gaussian mode for photodetection. An array of $N$ photodetectors is used to collect the power in each spatial mode. The (de)multiplexing process can be realized through diffraction or refraction optics. For instance, spatial light modulator (SLM) \cite{wang} and mode sorter \cite{huang2015mode} are usually employed in OAM-based FSO systems. Although some (de)multiplexing techniques can introduce additional power loss to the system, in this work, we assume that this process is near-perfect and no power loss is introduced as in \cite{Anguita:08}.  
	
	 Denote the field distribution of the spatial mode with mode state $k$ as $u_k(\mathbf{r},z)$ where $\mathbf{r}$ refers to the position vector and $z$ is the propagation distance. Note that $u_k(\mathbf{r},z)$ satisfies the orthonormality condition, i.e., \cite{Ren:14}
	 \begin{equation}
	 \int u_k(\mathbf{r},z) u_{k'}^*(\mathbf{r},z) \mathrm{d}\mathbf{r}=
	 \begin{cases}
	 1, & \text{if} \quad k=k'\\
	 0,& \text{if} \quad k\neq k'
	 \end{cases}.
	 \end{equation}
	 If a spatial mode with state $k$ is transmitted through the  atmospheric turbulence, the resulting wavefront on the receiver plane $\varphi_k(\mathbf{r},z)$ can be decomposed using the employed complete orthonormal spatial mode basis with specific coefficients as\cite{shengmei,Paterson}  
	 \begin{equation}
	 \varphi_k(\mathbf{r},z)=\sum\limits_{i=-\infty}^{+\infty} \alpha_{ki}u_i(\mathbf{r},z),
	 \end{equation}
	 where the $\alpha_{ki}$ refers to the coefficient between the transmitted mode state $k$ and the received mode state $i$ which can be obtained by the inner product 
	\begin{equation}
	\alpha_{ki}=\int\varphi_k(\mathbf{r},z)u_i^*(\mathbf{r},z) \mathrm{d} \mathbf{r}.
	\end{equation}
	{Note that in general $\alpha_{ki}$ is a complex value which is related to the instantaneous channel state\cite{Paterson}. The normalized power leaked from the state $k$ to the state $i$ after propagation through the atmosphere can be expressed by $|\alpha_{ki}|^2$ \cite{Anguita:08}. The statistical characteristics of $|\alpha_{ki}|^2$ which depends on the specific states $k$ and $i$ has been investigated in a few works. For instance, it is concluded that for OAM modes the self-channel fading, i.e., $|\alpha_{kk}|^2$, obeys Johnson $S_B$ distribution and the crosstalk fading, i.e., $|\alpha_{ki}|^2$ with $k\neq i$, on the other hand obeys exponential distribution \cite{Anguita_pdf}. {For statistically homogeneous and isotropic turbulence, the distribution of the random phase distortion is symmetric around the origin with a large variance, therefore it can be approximated as uniform distribution with high accuracy}\cite{lee_con}. {Similarly, in this paper the phase of the crosstalk fading $\alpha_{ki}$ denoted by} $\angle \alpha_{ki}$ {with $k\neq i$ is also assumed to be uniformly distributed within the interval $[0,2\pi]$. This approximation can be verified numerically under the turbulence conditions considered here. } 
	
	Denoting the transmitted mode set as $\mathcal{N
	}$, the combined transmitted optical field at the transmitter telescope can be expressed as $\sum_{k\in \mathcal{N
	}} \rho_ku_k(\mathbf{r},0)$ where $\rho_k$ is the modulated optical magnitude for the transmitted mode state $k$. We consider that $\rho_k$ obeys the average power constraint that $E[\rho_k^2]=P_t/N$ where $P_t$ is the totally transmitted average power and $N$ is the number of elements in the set $\mathcal{N}$. We assume that the transmitter does not have the channel state information so that the total power is uniformly allocated to all transmitted modes. In addition, the linewidth of the laser source is assumed to be narrow hence there is no significant relative temporal phase difference between the transmitted modes \cite{Legg}. The received optical field over the receiver telescope can then be written as 
	\begin{equation}
	\varphi(\mathbf{r},z)=\sum\limits_{i=-\infty}^{+\infty}\sum\limits_{k\in \mathcal{N
		}}^{}\rho_k\alpha_{ki}u_i(\mathbf{r},z),
	\end{equation}
	where $z$ is the propagation distance. After the mode demultiplexing, for the photodetector collecting the power in the mode state $i$, the received optical power is given by 
	\begin{align}\label{re_pow}
	y_i&=\int\bigg|\sum\limits_{k\in \mathcal{N
		}}^{}\rho_k\alpha_{ki}u_i(\mathbf{r},z)\bigg|^2\,d\mathbf{r}\\\nonumber
	&=\bigg|\sum\limits_{k\in \mathcal{N
		}}^{}\rho_k\alpha_{ki}\bigg|^2,
	\end{align}
	where the receiver aperture is considered big enough to collect all the power in the $i$th mode and the orthonormality of spatial modes is applied. Note that the effect of the ambient light, which is considered to be negligible compared to the crosstalk and thermal noise, is not included here. In (\ref{re_pow}) the signal and crosstalk are coherently superimposed, thus for the whole SMM system the vector of the received optical power $\mathbf{Y}=[y_1,\cdots,y_N]^T$ can be expressed as
	\begin{equation}
	\mathbf{Y}=|\mathbf{H}\boldsymbol{\rho}|^2,
	\end{equation}
	where $\boldsymbol{\rho}=[\rho_1,\cdots,\rho_N]^T$  is the vector of the transmitted signal and $\mathbf{H}$ is the channel matrix given by 
	\begin{equation}\label{H}
	\mathbf{H}=
	\begin{bmatrix}
	\alpha_{11} & \dots & \alpha_{N1}\\
	\vdots & \ddots &\vdots\\
	\alpha_{1N} & \dots & \alpha_{NN}\\
	\end{bmatrix}.
	\end{equation}
	This non-linear transformation between the received optical power $\mathbf{Y}$ and transmitted signal $\boldsymbol{\rho}$ is due to the square-law optical detection making the traditional MIMO-DSP techniques inapplicable to this system \cite{Arik:16,Yadin}. It is worth mentioning that unlike the system investigated here,  when mutually incoherent channels are considered (e.g., see \cite{Anguita:08}), the received optical power can be written as the incoherent superposition of the signal power and the crosstalk. Hence, the channel transformation is linear instead which is given by $\mathbf{Y}'=\mathbf{H}'\boldsymbol{\rho}'$, where $\boldsymbol{\rho}'=[\rho_1^2,\cdots,\rho_N^2]^T$ and 
	\begin{equation}\label{H'}
	\mathbf{H}'=
	\begin{bmatrix}
	|\alpha_{11}|^2 & \dots & |\alpha_{N1}|^2\\
	\vdots & \ddots &\vdots\\
	|\alpha_{1N}|^2 & \dots & |\alpha_{NN}|^2\\
	\end{bmatrix}.
	\end{equation}
	However, as explained before, such a mutually incoherent channel model is valid for FSO systems that may consume more spectral DOFs than required. We therefore focus on the mutually coherent channels as also considered in \cite{Arik:16,Yadin}.    
		
	Denoting the time of postdetection integration as $\tau$ which corresponds to the symbol duration, due to the effect of shot noise, the vector of the detected photon count can be modelled as a doubly stochastic Poisson process \cite{shane} with photon rate vector $\boldsymbol{\varLambda}=\mu\mathbf{Y}$ where the coefficient $\mu={\eta\tau}/{h\nu}$, $\eta$ is the quantum efficiency, $h$ is Plank's constant and $\nu$ is optical field frequency. Note that, in the literature, optical receivers are usually assumed to be either thermal noise or shot noise limited, however, here we consider a general scenario where both shot and thermal noise are taken into account \cite{pream}. Without loss of generality, we will focus on the multiplexed channel with mode state $i$ in the following derivation. The same analysis can be easily extended to other channels in the multiplexing system. Using (\ref{re_pow}), the photon rate $\varLambda_i$ for this channel can be rewritten by   
	\begin{equation}\label{mi}
	\varLambda_i=\mu\,\bigg|\rho_i\alpha_{ii}+\sum\limits_{k\in \mathcal{N
		},k\neq i} \rho_k\alpha_{ki}\bigg|^2,
	\end{equation}
	where $\rho_i\alpha_{ii}$ refers to the signal from intended spatial mode and the summation term is the interference from other channels. We assume that the receiver has the instantaneous channel state information (CSI) of the amplitude of the signal fading $|\alpha_{ii}|$, which can be easily estimated by exciting the mode state $i$ and collecting the received optical power in the same mode \cite{Bikhazi,Arik:16}. However, the instantaneous CSI of the interference fading is assumed to be unknown to the receiver and the receiver only has access to its statistical characteristics. {Based on the central limit theorem} \cite{Shah} {and the uniform distribution of} $\angle \alpha_{ki}$, with the increase of the number of transmitted modes, the interference term can be approximated as a narrowband complex Gaussian distributed noise with zero mean and variance $\sigma_{c,i}^2$ on each quadrature where 
	\begin{equation}\label{2}
	\sigma_{c,i}^2=\frac{P_t}{2N}\sum\limits_{k\in \mathcal{N
		},k\neq i} E[|\alpha_{ki}|^2].
	\end{equation}
	Note that the expectation of the crosstalk $|\alpha_{ki}|^2$ varies for different transmitted mode $k$ and received mode $i$ and can be measured at the beginning
	of the communication. The photon rate $\varLambda_i$ in (\ref{mi}) can thus be approximated as a non-central Chi square distributed random variable with PDF 
	\begin{equation}
	f_{\varLambda_i}(\varLambda_i)=\frac{1}{m_{c,i}}\mathrm{exp}\left(-\frac{\varLambda_i+m_{s,i}}{m_{c,i}}\right)I_0\left(\frac{2\sqrt{\varLambda_im_{s,i}}}{m_{c,i}}\right),
	\end{equation}  
	where 
	\begin{equation}
	m_{s,i}={\mu\rho_i^2\mid\alpha_{ii}\mid^2}
	\end{equation}
	is the average signal photon count,
	\begin{equation}\label{3}
	m_{c,i}={2\mu\sigma_{c,i}^2}
	\end{equation}
	refers to the average interference photon count and $I_0(\cdot)$ is the modified Bessel function with zero order. With this photon rate $\varLambda_i$, the probability of the detected photon count $n_i$ can be modelled as Laguerre distribution with PDF given by \cite{pream} 
	\begin{equation}
	f_{n_{i}}(n_{i})=\frac{m_{c,i}^{n_{i}}}{(1+m_{c,i})^{n_{i}+1}}\, \mathrm{exp}\left(-\frac{m_{s,i}}{1+m_{c,i}}\right)L_{n_{d}}\left(-\frac{m_{s,i}}{m_{c,i}(1+m_{c,i})}\right),
	\end{equation}
	where the Laguerre polynomial $L_{n}(x)=\sum_{j=0}^{x}C_n^j(-x)^j/j!$. 
	The characteristic function of this distribution can be expressed by 
	\begin{equation} \label{Cfunc}
	\Psi(j\omega)=\frac{1}{1+m_{c,i}(1-e^{j\omega})}\mathrm{exp}\left[\frac{-m_{s,i}(1-e^{j\omega})}{1+m_{c,i}(1-e^{j\omega})}\right]. 
	\end{equation}
	Based on this characteristic function, the mean and variance of $n_{i}$ are given by
	\begin{align}
	u_{i}&=m_{s,i}+m_{c,i},\\\nonumber
	\sigma_{i}^2&=u_{i}+m_{c,i}^2+2m_{s,i}m_{c,i}.
	\end{align}
	Note that in the expression of $\sigma_{i}^2$, $u_{i}$ is the shot noise introduced by Poisson photodetection process and $m_{c,i}^2+2m_{s,i}m_{c,i}$ results from the fluctuation of the rate $	\varLambda_i$ itself due to the randomness of the interference. If we further bring the thermal Gaussian noise into account, the output count can be expressed as 
	\begin{equation}
	n_{o,i}=n_{i}+n_{th,i},
	\end{equation}
	where $n_{th,i}$ is Gaussian noise with zero mean and variance $\sigma_{th}^2=2k_BT_o\tau/R_Lq^2$ \cite{pream}. Note that $k_B$ is the Boltzmann's constant, $R_L$ is the load resistance, $T_o$ is the receiver temperature in degrees Kelvin and $q$ is the electron charge. The mean and variance of $n_{o,i}$ can then be written as 
	\begin{align}\label{mean_var}
	u_{o,i}&=m_{s,i}+m_{c,i},\\\nonumber
	\sigma_{o,i}^2&=u_{o,i}+m_{c,i}^2+2m_{s,i}m_{c,i}+\sigma_{th}^2.
	\end{align}

	We would like to further emphasize that in this work the channel is modelled based on the photon counting statistics but the classical Poisson channel model which is commonly employed in optical communication systems cannot be applied. In fact, in most of the works applying photon counting analysis, the noise term in the rate of the doubly stochastic Poisson process is usually introduced by the ambient light with a bandwidth (optical bandwidth) much larger than the signal electrical bandwidth, as a result a large number of temporal modes of the noise is able to be detected, which allows the noise randomness to be averaged over all the temporal modes \cite{shane}. Therefore, the variation of the rate is smoothed out and the detected count with Laguerre distribution can be approximated by a Poisson distribution with high accuracy \cite{gagliardi}. However, in this work the noise term in the rate $\varLambda_i$ given in (\ref{mi}) is introduced by the crosstalk from other channels which has a bandwidth comparable to the signal electrical bandwidth. Therefore only one temporal mode is detected and the Laguerre count probability cannot be simplified to the classical Poisson probability.    
	\section{Performance Analysis}\label{perf}
	\subsection{Aggregate Achievable Rate}\label{AAR}
	The channel model considered in Section \ref{channel} is similar to that of the optical communication systems impaired by random background noise such as in systems employing optical preamplifiers \cite{Humblet,pream}. In order to proceed our analysis, the output photon counts $n_{o,i}$ can be approximated as a Gaussian distributed random variable with the mean and variance given by (\ref{mean_var}) as in \cite{Humblet,Li,pream}. After removing the bias introduced by the average interference photons $m_{c,i}$, the channel model can then be rewritten as 
	\begin{equation}\label{channelG}
	n_{o,i}=m_{s,i}+\sqrt{m_{s,i}}Z_{s,i}+Z_{0,i},
	\end{equation}
	where $Z_{s,i}$ and $Z_{0,i}$ are Gaussian distributed random variables with zero mean and variance
	\begin{equation}
	\sigma_{Z_{s,i}}^2=1+2m_{c,i},\quad
	\sigma_{Z_{0,i}}^2=m_{c,i}+m_{c,i}^2+\sigma_{th}^2,	
	\end{equation}
	respectively. The first term in (\ref{channelG}) refers to the signal, the second term is the signal/input-dependent noise which is introduced by the signal-induced shot noise and the fluctuation of the beat term in (\ref{mi}) due to the random interference, and the third term describes the signal-independent noise which is introduced by the shot noise caused by the interference, the fluctuation of the interference and the thermal noise. The exact expression for the capacity of such channel is unknown, however, its lower and upper bounds under input peak-power and average-power constraints have been investigated in \cite{Moser}. In this work, we are interested in the achievable rate (capacity lower bound) of the SMM systems with a total average-power constraint $P_t$. Using the achievable rate given by (23) in \cite{Moser}, for the channel with mode state $i$ in the SMM system, the achievable rate conditioned on the instantaneous signal fading $\alpha_{ii}$ can be expressed as 
	\begin{align}\label{rate}
\mathcal{C}_{i|\alpha_{ii}}=&\,\frac{1}{2}\mathrm{log}\frac{\mu|\alpha_{ii}|^2P_t}{N\sigma_{Z_{s,i}}^2}+\frac{1}{2}\mathrm{log}\left(1+\frac{2N\sigma_{Z_{s,i}}^2}{\mu|\alpha_{ii}|^2P_t}\right)-\frac{\mu|\alpha_{ii}|^2P_t}{N\sigma_{Z_{s,i}}^2}-1\\\nonumber
&+\frac{\sqrt{\!\mu|\alpha_{ii}|^2P_t\left(\mu|\alpha_{ii}|^2P_t+2N\sigma_{Z_{s,i}}^2\right)}}{N\sigma_{Z_{s,i}}^2}-\sqrt{\frac{\pi N\sigma_{Z_{0,i}}^2}{2\mu|\alpha_{ii}|^2P_t\sigma_{Z_{s,i}}^2}},
	\end{align}
which becomes tighter with the increase of the average transmitted power. The input to achieve this rate is half-normal distributed with PDF given by
\begin{equation}\label{rhopdf}
f_{\rho}(\rho)=\sqrt{\frac{2N}{\pi P_t}}\,\mathrm{exp}\left(-\frac{N\rho^2}{2P_t}\right).
\end{equation}
Since both of the noise variance $\sigma_{Z_{s,i}}^2$ and $\sigma_{Z_{0,i}}^2$ contain $m_{c,i}$ which depends on the transmitted power $P_t$ as shown in (\ref{2}) and (\ref{3}), it is expected that with the increase of $P_t$, the achievable rate will turn to be interference-limited and saturate at a fixed value. By substituting (\ref{2}) and (\ref{3}) into (\ref{rate}) and after some algebraic manipulations, {the asymptotic achievable rate at high $P_t$ can be achieved as}
\begin{equation}\label{asy}
\mathcal{C}^\mathrm{\infty}_{i|\alpha_{ii}}=\frac{1}{2} \mathrm{log}\left(\!\frac{1}{2}\gamma_i+2\!\right)-\!\frac{\gamma_i}{2}\!-\!1+\frac{\sqrt{\gamma_i(\gamma_i+4)}}{2}-\sqrt{\frac{\pi}{4\gamma_i}},
\end{equation}
\sloppy where $\gamma_i$ is the instantaneous asymptotic  signal-to-interference ratio (SIR) given by $\gamma_i=|\alpha_{ii}|^2/\sum_{k\in \mathcal{N},k\neq i} E[|\alpha_{ki}|^2]$.
Note that (\ref{asy}) is only related to the asymptotic SIR $\gamma_i$ which is related to the channel state and does not depend on the average transmitted power $P_t$. 
	
So far we have derived the instantaneous achievable rate for the multiplexed channel with mode state $i$ in the SMM system. When all channels in the system decode their data independently, the aggregate achievable rate (AAR) should be considered which is given by the summation of the achievable rates of $N$ channels, i.e.,  $\sum_{i\in\mathcal{N}}^{}\mathcal{C}_{i|\alpha_{ii}}$  \cite{Anguita:08}. In order to evaluate the overall performance of the system, the average AAR is employed as a performance metric which can be calculated by averaging over the channel states, i.e.,  
\begin{equation}\label{aver_cap}
\overline{\mathcal{C}}=E\left[\sum_{i\in\mathcal{N}}^{}\mathcal{C}_{i|\alpha_{ii}}\right].
\end{equation}
Considering the complicated achievable rate expression given in (\ref{rate}) and the fact that the complete statistical characteristics of the signal fading for different spatial modes are not available, an analytical solution for $\overline{\mathcal{C}}$ is intractable. In the next section we will numerically calculate $\overline{\mathcal{C}}$ by averaging over a large number of propagation instances generated by simulation of beam propagation using the random phase screen approach \cite{numerical}. Moreover, the average asymptotic AAR can be calculated using
\begin{equation}
\overline{\mathcal{C}}_\mathrm{\infty}=E\left[\sum_{i\in\mathcal{N}}^{}\mathcal{C}^\mathrm{\infty}_{i|\alpha_{ii}}\right].
\end{equation}


	\subsection{Numerical Results}\label{results}
	In this section, we present some simulation results for a typical SMM FSO system with mutually coherent channels, based on our analytical derivations in Section \ref{AAR}. For the numerical results, we focus on OAM orthogonal spatial mode set considering that it has attracted significant interest from scientific community recently \cite{willner}. However, we would like to emphasize that all the analytical derivations in this paper can also be applied to FSO systems employing other spatial modes such as HG and HB modes. 
	
	The optical field for OAM mode state $i$ at the transmitter plane is given by 
	\begin{equation}
	u_i(r,\phi,0)=\sqrt{\frac{2}{\pi|i|!}}\frac{1}{w_0}\left(\frac{\sqrt{2}r}{w_0}\right)^{|i|}L^i_0\left(\frac{2r^2}{w_0^2}\right)\mathrm{exp}\left(\frac{-r^2}{w_0^2}\right)\mathrm{exp}\left(-ji\phi\right),
	\end{equation} 
	where $w_0$ is the beamwidth for fundamental Gaussian beam at the transmitter plane, $L^i_0(\cdot)$ represents the generalized Laguerre polynomial and $r$ and $\phi$ refer to the radial distance and azimuthal angle, respectively. 
	\begin{figure}[!t]
		\centering
		\includegraphics[width=0.5\textwidth]{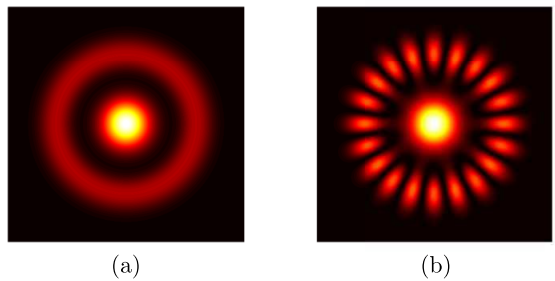}
		\caption{The intensity distribution for an OAM-based multiplexing system imaged at the transmitter plane where OAM mode set $\mathcal{N
			}=\{0, \pm10\}$ is employed, (a) mutually incoherent channels; (b) mutually coherent channels.}\label{mu_inco_co3}
	\end{figure}
	\figurename\textrm{ }\ref{mu_inco_co3} shows the difference of the intensity distribution imaged at the transmitter plane between OAM-based multiplexing employing mutually coherent and incoherent channels imaged at the transmitter plane. For mutually incoherent channels (\figurename\textrm{ }\ref{mu_inco_co3}(a)), the multiplexed beam intensity is simply the intensity superposition (incoherent addition) of transmitted modes. However, for mutually coherent channels (\figurename\textrm{ }\ref{mu_inco_co3}(b)), the optical fields are superimposed (coherent addition) and the multiplexed intensity pattern is more complicated due to the constructive and destructive interference between modes. Note that the coherent OAM mode superposition has also been investigated in \cite{Anguita:co} for high-dimensional modulation. 
	
	The propagation of the beams through atmosphere is numerically simulated using the split-step Fourier method \cite{numerical} and totally $5\times10^4$ propagation instances are simulated to ensure accurate simulation results. The propagation distance is set as $z=1$ km, the transmitted beam wavelength is $\lambda=850$ nm, the quantum efficiency is assumed equal to $\eta=1$, the receiver temperature $T_o=300$ K, the local resistance $R_L=50$ $\Omega$, the electrical bandwidth is $1$ GHz which corresponds to a symbol duration of $\tau=1$ ns and the beamwidth at the transmitter is $w_0=1.6$ cm which leads to the minimum beamwidth on the receiver plane \cite{Anguita:08}. In practical SMM systems, the range of spatial modes that can be employed is constrained by the limited transceiver sizes \cite{zhao}. In our simulation, the transceivers are designed so that OAM modes with state $-10$ to $+10$ can be transmitted and received successfully. Moreover, the inner and outer scales of the turbulence are assumed as $l_0=5$ mm and $L_0=20$ m, respectively. The phase screens are placed every $50$ m which are randomly generated based on the modified von Karman spectrum which is given by  
	\begin{equation}
	\Phi(\kappa)=\beta_1C_n^2\left[1+\beta_2(\kappa/\kappa_l)-\beta_3(\kappa/\kappa_l)^{7/6}\right]\frac{\mathrm{exp}\left(-\kappa^2/\kappa_l^2\right)}{(\kappa_0^2+\kappa^2)^{11/6}},
	\end{equation}   
	where $\beta_1=0.033$, $\beta_2=1.802$, $\beta_3=0.254$, $\kappa_l=3.3/l_0$, $\kappa_0=2\pi/L_0$ and $C_n^2$ is the refractive index structure constant. In the simulation, we choose two values for $C_n^2$, i.e., $1\times10^{-15} \,\mathrm{m}^{-2/3}$ and $6\times10^{-15} \,\mathrm{m}^{-2/3}$. According to the definition of Rytov variance $\sigma_R^2=1.23C_n^2k^{7/6}z^{11/6}$ where $k=2\pi/\lambda$, these two $C_n^2$ values correspond to $\sigma_R^2=0.04$ and $\sigma_R^2=0.24$, respectively.

\begin{figure}[!t]
	\centering
	\includegraphics[width=0.6\textwidth]{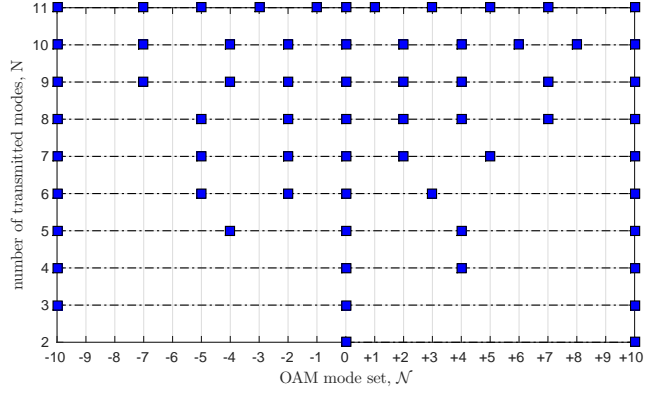}
	\caption{For different number of transmitted spatial modes $N$ in SMM system, the optimal set of transmitted modes $\mathcal{N}$ which maximize the average asymptotic AAR $\overline{\mathcal{C}}_\mathrm{\infty}$ when $C_n^2=1\times10^{-15} \,\mathrm{m}^{-2/3}$.}\label{optimal_modes}
\end{figure} 

In FSO SMM systems, the selection of the transmitted mode set $\mathcal{N}$ is essential because of the different crosstalk characteristics of the spatial modes when propagate through the atmosphere. In this work, the OAM modes that can be employed for transmission are ranged from $-10$ to $+10$ and we are interested in the optimal set $\mathcal{N}$ that can maximize the average asymptotic AAR $\overline{\mathcal{C}}_\mathrm{\infty}$ under different turbulence conditions. Note that for each channel in the transmitted mode set, the instantaneous asymptotic achievable rate can be calculated using (\ref{asy}). \figurename\textrm{ }\ref{optimal_modes} plots the optimal transmitted mode set $\mathcal{N}$ with respect to the number of elements $N$ when $C_n^2=1\times10^{-15} \,\mathrm{m}^{-2/3}$ by using exhaustive search. Note that for other turbulence conditions, similar optimal mode sets can be observed. One can see that the fundamental Gaussian beam with OAM mode $i=0$ is always preferable for different $N$, because this mode has the best ability of keeping the original mode status after propagating through atmosphere \cite{Anguita:08}. It is also shown that the relative separations of the transmitted mode states should be chosen as large as possible. For example, for three-mode transmission $N=3$, the optimal mode set is $\mathcal{N}=\{0,\pm10\}$ and for $N=5$, the optimal set is $\mathcal{N}=\{0,\pm4,\pm10\}$. This is because at high $P_t$ regime, the multiplexing systems are interference-limited and those systems with larger mode separation, which indicates smaller crosstalk between channels and hence larger asymptotic SIRs, can achieve higher AAR. Note that similar phenomenon is also observed for SMM systems with mutually incoherent channels \cite{Anguita:08}. Furthermore, one can also observe from \figurename\textrm{ }\ref{optimal_modes} that when $N$ is odd number, the mode set $\mathcal{N}$ is always symmetrical around the OAM state $0$. 
 
\begin{figure}[!t]
	\centering
	\includegraphics[width=0.6\textwidth]{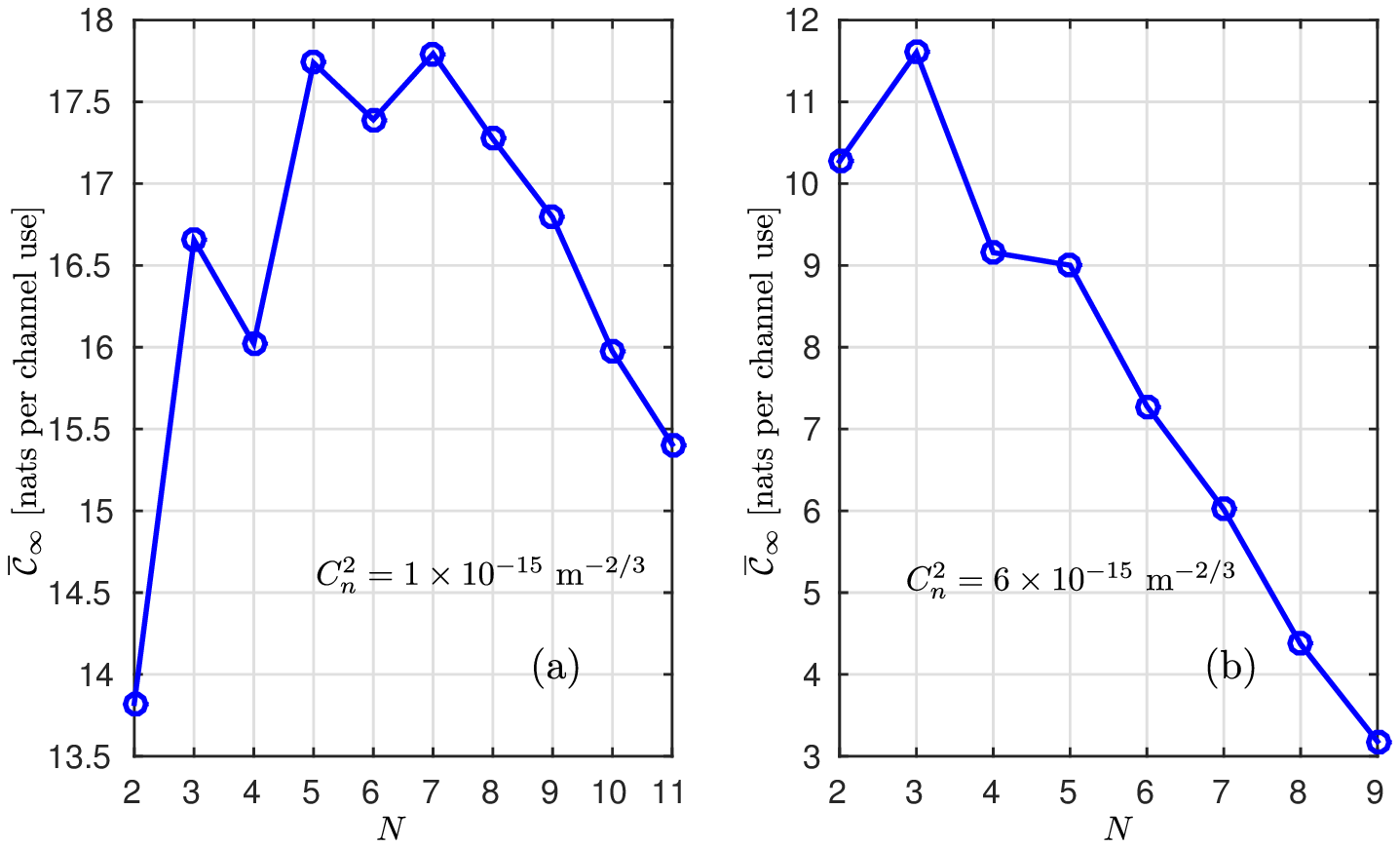}
	\caption{The average asymptotic AAR $\overline{\mathcal{C}}_\mathrm{\infty}$ versus the number of transmitted modes $N$ under different turbulence conditions. (a) $C_n^2=1\times10^{-15} \,\mathrm{m}^{-2/3}$; (b) $C_n^2=6\times10^{-15} \,\mathrm{m}^{-2/3}$.}\label{max_asy_N}
\end{figure} 

The average asymptotic AAR $\overline{\mathcal{C}}_\mathrm{\infty}$ versus $N$ under different turbulence conditions is plotted in \figurename\textrm{ }\ref{max_asy_N}. Note that for each $N$, the optimal set $\mathcal{N}$ is used according to \figurename\textrm{ }\ref{optimal_modes}. One can see that with the increase of $N$, $\overline{\mathcal{C}}_\mathrm{\infty}$ firstly increases and then decreases. This is because when $N$ is small, the SMM system benefits from the additional spatial DOFs explored by adding more transmitted modes or channels, hence higher $\overline{\mathcal{C}}_\mathrm{\infty}$ can be achieved with the increase of $N$. Note that the initial increase of the $\overline{\mathcal{C}}_\mathrm{\infty}$ might not be monotonically with respect to $N$. For instance when $C_n^2=1\times10^{-15} \,\mathrm{m}^{-2/3}$, the $\overline{\mathcal{C}}_\mathrm{\infty}$ when $N=4$ is even smaller than that of $N=3$. This is due to the symmetry and asymmetry of $\mathcal{N}$ with respect to the mode state $0$ when $N$ is odd and even, respectively. Actually, \figurename\textrm{ }\ref{max_asy_N} indicates that for small $N$, the mode sets $\mathcal{N}$ with odd elements are more preferable than those with even elements. On the other hand, adding more transmitted modes also introduces additional crosstalk to other channels, which degrades the performance of other channels. Therefore with the further increase of $N$, the increase of $\overline{\mathcal{C}}_\mathrm{\infty}$ due to the additional DOF might not be able to compensate the additional degradation introduced, which in turn results in the decrease of $\overline{\mathcal{C}}_\mathrm{\infty}$. As a result, an optimal $N$ exists which can achieve the maximal $\overline{\mathcal{C}}_\mathrm{\infty}$. For instance, when $C_n^2=1\times10^{-15} \,\mathrm{m}^{-2/3}$ and $C_n^2=6\times10^{-15} \,\mathrm{m}^{-2/3}$, the optimal number of transmitted modes are $N=7$ and $N=3$ which correspond to the mode sets $\mathcal{N}=\{0,\pm2,\pm5,\pm10\}$ and $\mathcal{N}=\{0,\pm10\}$, respectively. Note that for stronger turbulence, the optimal number of channels significantly decreases because of the stronger crosstalk effects. The results shown in \figurename\textrm{ }\ref{optimal_modes} and \figurename\textrm{ }\ref{max_asy_N} are valuable for the design of the practical FSO SMM systems. Since the transceiver sizes in practical SMM systems limit the range of spatial modes that can be employed in the system, using the above figures one can select the optimal transmitted mode set $\mathcal{N}$, which is associated with the turbulence condition, to maximize the average asymptotic AAR. 
\begin{figure}[!t]
	\centering
	\includegraphics[width=0.7\textwidth]{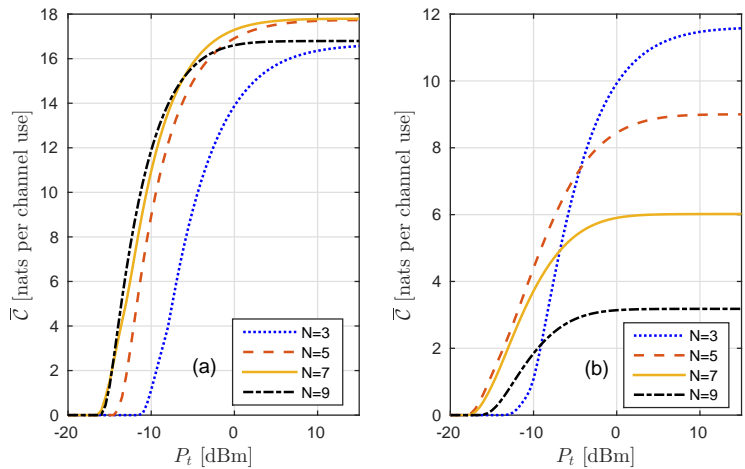}
	\caption{The average AAR $\overline{\mathcal{C}}$ versus the average transmitted power $P_t$ for different number of transmitted modes (a) $C_n^2=1\times10^{-15} \,\mathrm{m}^{-2/3}$; (b) $C_n^2=6\times10^{-15} \,\mathrm{m}^{-2/3}$. }\label{cap_with_SNR}
\end{figure} 

The average AAR $\overline{\mathcal{C}}$ given by (\ref{aver_cap}) with respect to $P_t$ for different $N$ is plotted in \figurename\textrm{ }\ref{cap_with_SNR}. Note that still for each $N$ the optimal set $\mathcal{N}$ which results in the maximal $\overline{\mathcal{C}}_\mathrm{\infty}$ is chosen according to \figurename\textrm{ }\ref{optimal_modes}. In lower $P_t$ regime, with the increase of $P_t$, $\overline{\mathcal{C}}$ usually grows much faster for the systems with larger $N$ than those with smaller $N$ due to the more spatial DOFs they explored. For instance, by increasing $P_t$ from $-15$ dBm to $-10$ dBm, an increase of $8.9$ nats per channel use can be observed for $N=5$ when $C_n^2=1\times10^{-15} \,\mathrm{m}^{-2/3}$. However, the corresponding increments for $N=7$ and $N=9$ are $9.4$ and $10.6$ nats per channel use, respectively. In high $P_t$ regime, the system turns to be interference-limited and $\overline{\mathcal{C}}$ saturates at a fixed value, i.e., $\overline{\mathcal{C}}_\mathrm{\infty}$. As mentioned before, an optimal number of channels exists which can achieve the maximal $\overline{\mathcal{C}}_\mathrm{\infty}$. For instance, when $C_n^2=1\times10^{-15} \,\mathrm{m}^{-2/3}$, $N=7$ is the number of the transmitted modes which maximizes $\overline{\mathcal{C}}$. In \figurename\textrm{ }\ref{cap_with_SNR}(a), one can see that the asymptotic rate for $N=3$ is $16.6$ nats per channel use. By increasing $N$ to $7$, the corresponding rate increases to $17.8$ nats per channel use. However, further increasing $N$ to $9$ in turn decreases the asymptotic rate which results in the rate $16.8$ nats per channel use. Similar phenomena can also be observed for stronger turbulence $C_n^2=6\times10^{-15} \,\mathrm{m}^{-2/3}$ in \figurename\textrm{ }\ref{cap_with_SNR}(b), however, with this turbulence condition the optimal $N$ is only $3$ and further increasing $N$ will decrease the asymptotic rate at high $P_t$. In addition, \figurename\textrm{ }\ref{cap_with_SNR} also indicates that in case of operation at lower $P_t$ regime the optimum number of modes will increase from that of the high $P_t$ case.     

\section{SMM with Mode Diversity} \label{MD}
\subsection{Mode Diversity}	
Although mode-multiplexing can significantly increase the aggregated capacity of the FSO systems, the reliability of each multiplexed channel might be strongly impaired by the turbulence. Therefore some techniques have to be employed to suppress the effect of crosstalk and improve the communication reliability. When coherent detection is employed, MIMO-DSP is commonly employed to mitigate interference effects\cite{Ren:16}. However it cannot be applied in IM/DD SMM systems with mutually coherent channels considered here due to the non-linear channel transformation. Another method that can be employed is the adaptive optics \cite{Ren:14,shengmei}, which might be too  expensive to be used in practical commercial cost-effective FSO links. In our previous work, zero-forcing beamforming has been investigated in such systems \cite{shenjie}. In this paper, we propose to use a \textit{mode diversity} scheme to improve the reliability of the SMM channels, which is easy to implement in practice and is able to significantly improve the outage performance. We would like to emphasize that although here we consider SMM FSO systems with mutually coherent channels, mode diversity can also be used in those systems with mutually incoherent channels \cite{Mehrpoor}.
		\begin{figure}[!t]
			\centering
			\includegraphics[width=0.55\textwidth]{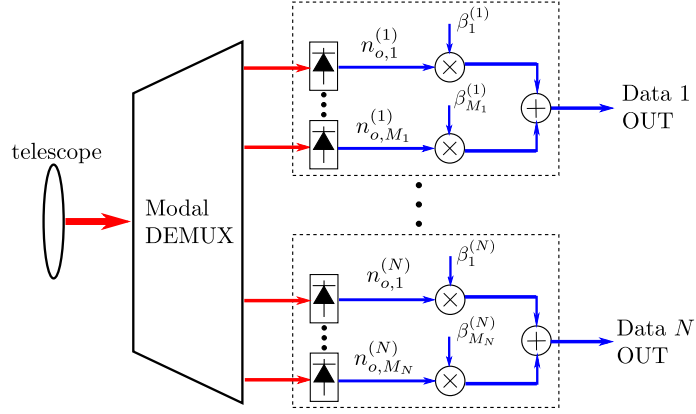}
			\caption{The receiver of FSO SMM system with mode diversity for mutually coherent channels. }\label{div_scheme}
		\end{figure} 

As mentioned in Section \ref{channel} that after propagation through the atmosphere, the power of the transmitted modes will leak to other spatial modes. Take OAM mode propagation as an example, it is concluded that the power in the intended mode is more likely to leak to those OAM modes with adjacent mode states and this leakage becomes stronger with the increase of the transmitted OAM mode state \cite{Anguita:08,willner}. In traditional direct-detection SMM, only the power in those transmitted modes are detected as shown in \figurename\textrm{ }\ref{scheme} thereby the SMM system can be described by $N$ multiplexed SISO channels. However, due to the turbulence-induced power leaking, the received power in modes other than the ones employed for multiplexing might also contain considerable signal power and hence can be used to improve the reliability of the channels by the means of diversity. The schematic of the proposed receiver with mode diversity is plotted in \figurename\textrm{ }\ref{div_scheme}. After receiving the incoming optical field, modal demultiplexing is applied. However, not only the optical power in those modes within the transmitted mode set $\mathcal{N}$ is detected, the optical signals in some other modes are also detected by the photodetector array. The detected optical signals are then combined together to realize the diversity. With this receiver scheme, the previous $N$ SISO links in the multiplexing system turn into $N$ SIMO links each with receive diversity. For instance, for the channel operated on mode state $i$, denoting the mode set for diversity as $\mathcal{M}_i$ with $M_i$ elements in it, the detected signals in these modes act as diversity branches and are combined after multiplying by distinct coefficients $\beta_{j}^{(i)}$ with $j\in[1,2,\cdots,M_i]$. It is worth mentioning that in practical SMM systems, one can actually easily get access to the received signals in numerous spatial modes with small power loss and no additional hardware complexity by using some well-designed optical devices such as the mode sorter for OAM-based SMM systems   \cite{huang2015mode}. When mode sorter is employed, the received optical signals in different spatial modes are transformed into laterally separated and elongated spots, therefore the received signal in any spatial mode supported by the receive aperture can be collected at different elements of an already employed integrated detector array. 

Taking the channel with the transmitted mode state $i$ in the SMM system as an example, the detected photon counts in the presence of mode diversity can be expressed as 
\begin{equation}\label{noi}
\tilde{n}_{o,i}=\sum_{j\in\mathcal{M}_i}\beta_j^{(i)} n_{o,j}^{(i)},
\end{equation}
where $\beta_j^{(i)}$ is the weighting coefficients and $n_{o,j}^{(i)}$ is the photon counts in the combining branch with mode state $j$. Invoking (\ref{channelG}), $n_{o,j}^{(i)}$ can be written as
\begin{equation}\label{noj}
n_{o,j}^{(i)}=\mu\rho_i^2|\alpha_{ij}|^2+\sqrt{\mu\rho_i^2|\alpha_{ij}|^2}Z_{s,j}^{(i)}+Z_{0,j}^{(i)},
\end{equation}
where $Z_{s,j}^{(i)}$ and $Z_{0,j}^{(i)}$ are still zero-mean Gaussian random variable with variance
\begin{align}\label{var01}
\sigma_{Z_{s,j}^{(i)}}^2&=1+2m_{c,j}^{(i)},\\\nonumber
\sigma_{Z_{0,j}^{(i)}}^2&=m_{c,j}^{(i)}+\left(m_{c,j}^{(i)}\right)^2+\sigma_{th}^2,
\end{align}
where $m_{c,j}^{(i)}$ is the average crosstalk photon count introduced by other multiplexed channels given by
\begin{equation}\label{mcij}
m_{c,j}^{(i)}=\frac{\mu P_t}{N}\sum_{k\in\mathcal{N},k\neq i}E[|\alpha_{kj}|^2].
\end{equation}
substituting (\ref{noj}) into (\ref{noi}), one can rewrite the output of the combiner as
\begin{equation}\label{noitil}
\tilde{n}_{o,i}=\mu\rho_i^2\sum_{j\in\mathcal{M}_i}\beta_j^{(i)} |\alpha_{ij}|^2+\sqrt{\mu\rho_i^2}\sum_{j\in\mathcal{M}_i}\beta_j^{(i)}|\alpha_{ij}|Z_{s,j}^{(i)}+\sum_{j\in\mathcal{M}_i}\beta_j^{(i)}Z_{0,j}^{(i)},
\end{equation}
where as (\ref{channelG}) the first term is the signal, the second term is the signal dependent noise and the third therm is the signal independent noise. The signal-to-interference-noise ratio (SINR) of the instantaneous output of the combiner conditioned on the channel fadings $|\alpha_{ij}|^2$ can then be expressed as \cite{gagliardi}
\begin{equation}\label{SNR2}
\zeta_i=\frac{\mu P_t\left(\sum_{j\in\mathcal{M}_i}\beta_j^{(i)} |\alpha_{ij}|^2\right)^2}{N \sum_{j\in\mathcal{M}_i}\left(\beta_j^{(i)}\right)^2\left[|\alpha_{ij}|^2\sigma_{Z_{s,j}^{(i)}}^2+\frac{N}{\mu P_t}\sigma_{Z_{0,j}^{(i)}}^2\right]},
\end{equation} 
where the average transmitted power constraint $E[\rho_i^2]=P_t/N$ is applied. Now we consider the choice of the weighting coefficient $\beta_j^{(i)}$. When all the coefficient is set as unity, the so-called equal gain combining (EGC) is realized \cite{Shah}. EGC is attractive due to its ease of implementation in practice. A more advanced choice of the coefficients that can maximize the SINR can also be employed here. This optimal combining is called the maximal ratio combining (MRC) and we will focus on this combining method in the following discussion. According to the  Cauchy-Schwarz inequality, the summation in the numerator of (\ref{SNR2}) satisfies
  \begin{equation*}
\left(\sum_{j\in\mathcal{M}_i}\beta_j^{(i)} |\alpha_{ij}|^2\right)^2\leq \sum_{j\in\mathcal{M}_i}\left(\beta_j^{(i)}\right)^2\left[|\alpha_{ij}|^2\sigma_{Z_{s,j}^{(i)}}^2+\frac{N}{\mu P_t}\sigma_{Z_{0,j}^{(i)}}^2\right]\times \sum_{j\in\mathcal{M}_i}\frac{|\alpha_{ij}|^4}{|\alpha_{ij}|^2\sigma_{Z_{s,j}^{(i)}}^2+\frac{N}{\mu P_t}\sigma_{Z_{0,j}^{(i)}}^2},\nonumber
  \end{equation*}
where the equality holds when 
\begin{equation}\label{beta}
\beta_j^{(i)}=\upsilon\,\frac{|\alpha_{ij}|^2}{|\alpha_{ij}|^2\sigma_{Z_{s,j}^{(i)}}^2+\frac{N}{\mu P_t}\sigma_{Z_{0,j}^{(i)}}^2},
\end{equation}
and $\upsilon$ is an arbitrary constant. Equation (\ref{beta}) gives the expression of the coefficients for MRC which results in the maximal output SINR. Note that different from the MRC in AWGN channel where the optimal coefficient is simply the fading gain (for real-valued fading) \cite{pream,Shah}, the optimal coefficient here is related not only to the fading but also to the transmitted signal power $P_t$. This is due to the fact that the investigated channel contains signal-dependent noise as illustrated in (\ref{noj}). Substituting (\ref{beta}) into (\ref{SNR2}), one can get the maximal SINR as
\begin{equation}\label{SNRMRC}
\zeta_i=\frac{\mu P_t}{N}\sum_{j\in\mathcal{M}_i}\frac{|\alpha_{ij}|^4}{|\alpha_{ij}|^2\sigma_{Z_{s,j}^{(i)}}^2+\frac{N}{\mu P_t}\sigma_{Z_{0,j}^{(i)}}^2},
\end{equation}
which can be regarded as the summation of SINRs of all diversity branches. Considering the expressions of the variance $\sigma_{Z_{0,j}^{(i)}}^2$ and $\sigma_{Z_{s,j}^{(i)}}^2$ given in (\ref{var01}), the asymptotic SINR at high $P_t$ can be written as
\begin{equation}\label{SNRasy}
\zeta_i^\mathrm{\infty}=\sum_{j\in\mathcal{M}_i}\frac{|\alpha_{ij}|^4}{2|\alpha_{ij}|^2\sum_{k\in\mathcal{N},k\neq i}E[|\alpha_{kj}|^2]+\left(\sum_{k\in\mathcal{N},k\neq i}E[|\alpha_{kj}|^2]\right)^2},
\end{equation}
which is not signal power dependent any more as expected.
		
\subsection{Diversity Mode Set}	\label{optdiv}	
So far, we have derived the coefficients for mode diversity with MRC combining. In this section, we consider the selection of the mode diversity set for each multiplexed channel, i.e., $\mathcal{M}_i$, which is essential and is directly associated with the reliability improvement. The performance of MRC combining always benefits from adding more branches, because it is able to adjust the combining coefficients given in (\ref{beta}), so that the output SINR is the summation of the branch SINRs. Therefore, the best diversity mode set for each multiplex channel should include the whole mode states that can be detected. However, increasing the number of combining branches will definitely make the receiver design as well as the channel estimation process more complicated. Furthermore, the SINRs of some branches might be very small and make little contributions to the enhancement of the output SINR. Thus it is valuable to find out the diversity mode set with least number of branches which achieves relatively high output SINR.         
 
It is known that with the decrease of the correlation of the fadings met by different branches, better diversity performance can be achieved \cite{Navidpour}. Therefore the correlation between the fadings of combining branches, i.e., $|\alpha_{ij}|^2$, should be considered. Different from the traditional diversity systems where distinct branches have identical SINR statistical characteristics, the branches in the mode diversity are inherently different because both the statistics of the signal fading $|\alpha_{ij}|^2$ and the values of the noise variance vary with the received mode state $j$. Among all of the modes that can be employed for diversity, the received signal in the mode with the same state as the transmitted mode, i.e., $j=i$, is obviously the one with highest average SINR, considering that the power conserved in the intended mode, i.e., $|\alpha_{ii}|^2$, is usually much larger than that leaks to other modes, i.e., $|\alpha_{ij}|^2$ with $j\neq i$. Thus the branch with $j=i$ is the most preferable branch and can be treated as the dominant one in the proposed SIMO link. Hence the correlation coefficients between $|\alpha_{ii}|^2$ and the fadings of the other branches $|\alpha_{ij}|^2$ are important. 

		\begin{figure}[!t]
			\centering
			\includegraphics[width=0.55\textwidth]{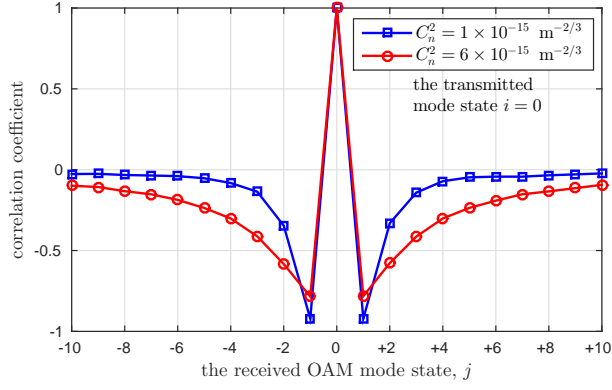}
			\caption{The correlation coefficient between the fading in the dominant branch $|\alpha_{ii}|^2$ and $|\alpha_{ij}|^2$ where the transmitted mode is $i=0$. }\label{correlation}
		\end{figure} 
To see this correlation relationship more clearly, we take the channel with $i=0$ in the OAM-based SMM system as an example. The correlation coefficient between $|\alpha_{ii}|^2$ and $|\alpha_{ij}|^2$ is plotted in \figurename\textrm{ }\ref{correlation} where the transmitted mode state is $i=0$. One can see that those received modes with states closer to the transmitted mode have high inverse correlation. For instance, the correlation between $|\alpha_{00}|^2$ and $|\alpha_{0+1}|^2$ is $-0.92$ when $C_n^2=1\times10^{-15}$ $\mathrm{m}^{-2/3}$, however, with the increase of the mode state difference between the transmitted and received modes, the correlation coefficients increase and approach zero. This is an expected result, because the total transmitted power is conserved and when the power remained in the intended mode is low, the transmitted power will  more likely be leaked to those modes with adjacent mode states, which results in high signal power in adjacent modes. As a result, the signal power in adjacent modes is negatively correlated to the power reserved in the transmitted mode \cite{Anguita_pdf}. This correlation relationship will spread more when the turbulence becomes stronger as shown in \figurename\textrm{ }\ref{correlation} due to the leakage of the power to more adjacent modes. We would like to emphasize that similar correlation relationship can also be observed when other spatial modes are transmitted. Based on the above discussion, one can conclude that the received signals in those modes with states closer to the transmitted mode state are more preferable to be used for mode diversity because of the highly negative correlation with the fading of the dominant branch. 
On the other hand, for each multiplexed channel in the SMM system, when the mode state of a branch is closer to that of other multiplexed channels, the power of the crosstalk contained in that branch increases. As a result, both noise variances $\sigma_{Z_{0,j}^{(i)}}^2$ and $\sigma_{Z_{s,j}^{(i)}}^2$ increase and according to (\ref{SNRMRC}) the SINR of this branch decreases and approaches zero. Therefore, it is expected that branches with mode states close to the transmitted mode state are preferred for mode diversity not only because of higher diversity gain (due to negative correlation) but also high power gain (due to less crosstalk).  
		\begin{figure}[!t]
			\centering
			\includegraphics[width=0.75\textwidth]{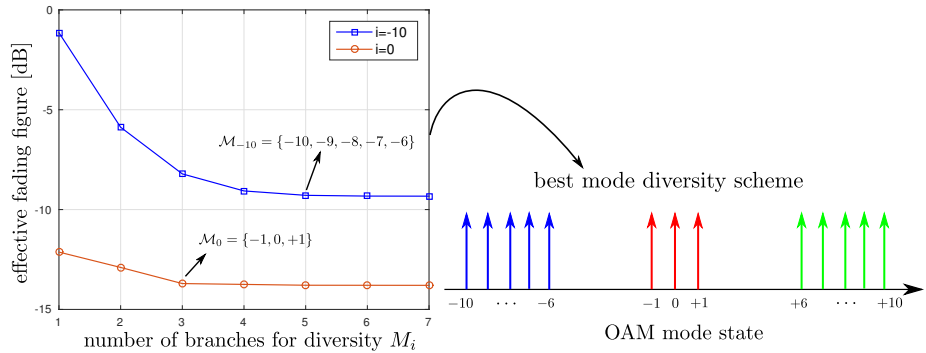}
			\caption{The EFF versus the number of branches for diversity, i.e., $M_i$, for different transmitted mode states $i$; $C_n^2=6\times10^{-15} \,\mathrm{m}^{-2/3}$ and the transmitted mode set is $\mathcal{N}=\{0,\pm10\}$.    }\label{EFF6E15}
		\end{figure} 
		
Now we would like to justify our expectation using numerical simulations. 
In order to measure the diversity performance properly, we employ the effective fading figure (EFF)  which can quantify the severity of the fading and the effectiveness of diversity systems on reducing signal fluctuations \cite{shin}. EFF is defined as the variance-to-mean-square ratio of the instantaneous combiner output SINR as  
\begin{equation}
EFF\,\left(\mathrm{dB}\right)=10\,\mathrm{log}_{10}\left\{\frac{\mathrm{Var}[\zeta_i]}{\left(E[\zeta_i]\right)^2}\right\},
\end{equation}
{where $\mathrm{Var}[\cdot]$ refers to the variance of the random variable. Note that the definition of EFF is close to the concept of the amount of fading (AF) which is commonly used in literature to assess the severity of the fading met at the output of a single fading channel} \cite{Hasna,Charash}. 

		\begin{figure}[!t]
			\centering
			\includegraphics[width=0.75\textwidth]{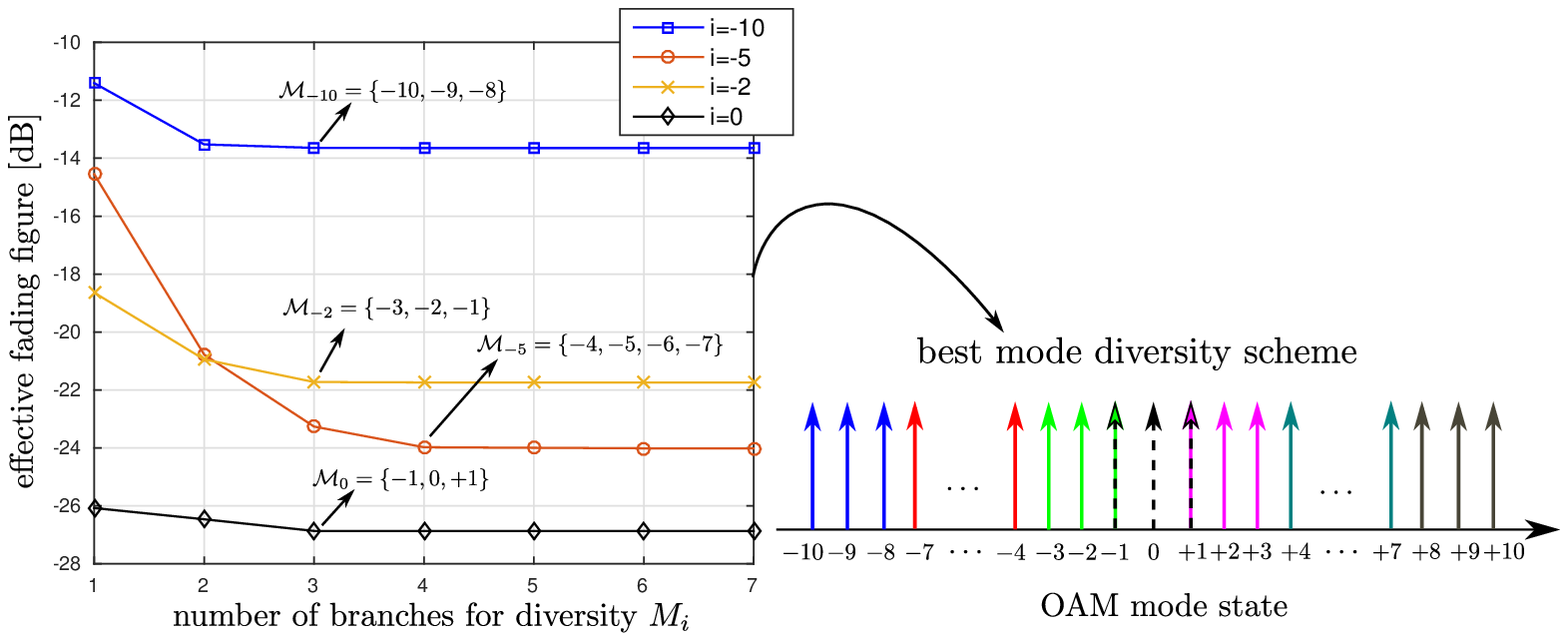}
			\caption{The EFF versus the number of received modes for diversity, i.e., $M_i$ for different transmitted mode states $i$; $C_n^2=1\times10^{-15} \,\mathrm{m}^{-2/3}$ and the transmitted mode set is $\mathcal{N}=\{0,\pm2,\pm5,\pm10\}$.    }\label{EFF1E15}
		\end{figure} 

For OAM-based SMM systems, the EFF versus the number of branches for diversity $M_i$ is plotted in \figurename\textrm{ }\ref{EFF6E15} and \ref{EFF1E15} for $C_n^2=6\times10^{-15} \,\mathrm{m}^{-2/3}$ and $C_n^2=1\times10^{-15} \,\mathrm{m}^{-2/3}$, respectively. Note that as in Section \ref{perf} we still focus on the high transmitted power regime, hence the expression of SINR is given by (\ref{SNRasy}). In addition, the transmitted mode set which maximizes the average asymptotic AAR is chosen. Since the employed $\mathcal{N}$ is symmetrical with respect to state $0$, the EFF of the channels with positive states $i$ are not plotted in the figures for simplicity, which are the same as the channels with corresponding {negative states}. Furthermore, for each $M_i$ we plot the EFF of a diversity mode set $\mathcal{M}_i$ which minimizes EFF through exhaustive search. From \figurename\textrm{ }\ref{EFF6E15} and \ref{EFF1E15} one can see that for every multiplexed channel with the increase of ${M}_i$, the EFF firstly decreases and then saturates on a fixed value. This justifies our expectation that adding branches is beneficial to the diversity system, however, with the increase of branches, the improvement of the diversity performance turns to be negligible. In these figures, we also point out the diversity mode sets with the least elements when the EFFs are saturated. we denote these mode sets as the best mode sets for diversity in the sense that they can achieve the best diversity performance with simplest receiver design. One can also see that the elements in the best mode set are all close to the transmitted mode state. For instance, in \figurename\textrm{ }\ref{EFF6E15} when $C_n^2=6\times10^{-15} \,\mathrm{m}^{-2/3}$ and $\mathcal{N}=\{0，\pm10\}$, the EFF is $-1.14$ dB for the channel with mode $i=-10$ in the absence of diversity. With the increase of the number of modes for diversity, EFF decreases and approaches to a fixed value $-9.3$ dB. The best diversity mode set is given by $\mathcal{M}_{-10}=\left\{-10,-9,-8,-7,-6\right\}$ and further increase in the number of combining branches can not improve the diversity gain. Note that the multiplexing channel with $i=+10$ (which is not plotted in \figurename\textrm{ }\ref{EFF6E15}) has symmetrical best mode set as $i=-10$, i.e., $\mathcal{M}_{+10}=\left\{+10,+9,+8,+7,+6\right\}$. Moreover, one can also determine the optimal diversity mode set for the channel with $i=0$ as $\mathcal{M}_{0}=\left\{0,\pm1\right\}$. Note that according to the simulation, although the channel with $i=-10$  benefits more from the mode diversity with larger EFF reduction, its minimal EFF is still larger than that with $i=0$. This is because the channel with $i=0$ is inherently superior to the channel with $i=-10$ \cite{Anguita:08}. Similar results can also be observed in \figurename\textrm{ }\ref{EFF1E15} when $C_n^2=1\times10^{-15} \,\mathrm{m}^{-2/3}$ with the transmitted mode set $\mathcal{N}=\{0,\pm2,\pm5,\pm10\}$, where the best diversity mode sets are $\mathcal{M}_{-10}=\{-10,-9,-8\}$, $\mathcal{M}_{-5}=\{-4,-5,-6,-7\}$, $\mathcal{M}_{-2}=\{-3,-2,-1\}$,  and $\mathcal{M}_{0}=\{0,\pm1\}$.

\subsection{Outage probability}
		\begin{figure}[!t]
			\centering
			\includegraphics[width=0.55\textwidth]{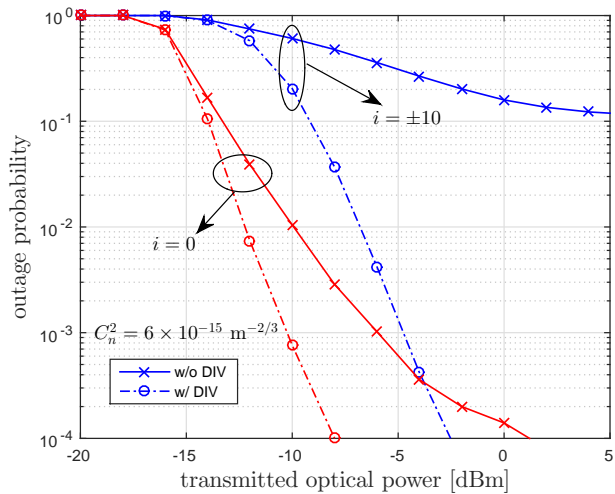}
			\caption{{The outage probability} versus the transmitted optical power for OAM-based SMM system with and without the mode diversity where $\mathcal{N}=\{0,\pm10\}$ and $C_n^2=6\times10^{-15} \,\mathrm{m}^{-2/3}$. DIV: mode diversity.    }\label{SINR_pow3}
		\end{figure} 

In order to evaluate the reliability improvement provided by the mode diversity, the outage probability will be investigated in this section. Outage probability is commonly employed in high-speed FSO communication systems to evaluate the reliability of the link, due to the slow-varying property of the atmospheric turbulence \cite{relay}. It is defined as the probability when the SINR is failing to achieve a prescribed threshold $\zeta_{th}$ and can be expressed as
\begin{equation}
P_\mathrm{out}=\mathrm{Pr}\left\{\zeta_i<\zeta_\mathrm{th}\right\}.
\end{equation}
		\begin{figure}[!t]
			\centering
			\includegraphics[width=0.55\textwidth]{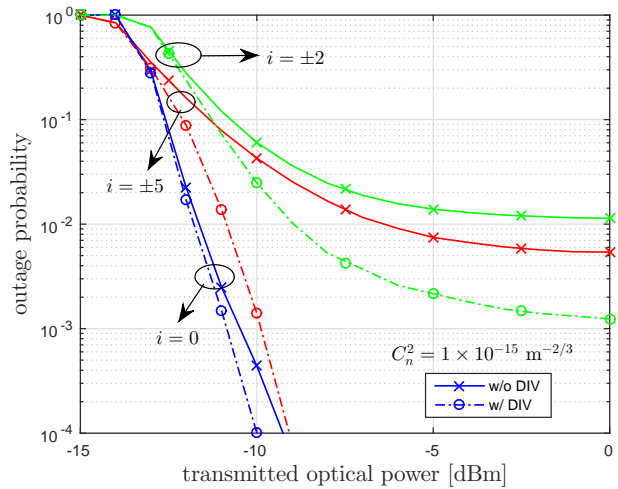}
			\caption{The outage probability versus the transmitted optical power for OAM-based SMM system with and without the mode diversity where $\mathcal{N}=\{0,\pm2,\pm5,\pm10\}$ and $C_n^2=1\times10^{-15} \,\mathrm{m}^{-2/3}$. The outage probability for the channel with $i=\pm10$ is omitted for the sake of clarity. DIV: mode diversity.    }\label{SINR_pow7}
		\end{figure} 
The outage probability versus the transmitted optical power with and without the mode diversity is plotted in \figurename\textrm{ }\ref{SINR_pow3} and \ref{SINR_pow7} for $C_n^2=6\times10^{-15} \,\mathrm{m}^{-2/3}$ and $C_n^2=1\times10^{-15} \,\mathrm{m}^{-2/3}$, respectively. For each multiplexed channel, the best mode set for diversity as discussed in Section \ref{optdiv} is employed. From these two figures one can observe the significant improvement of the outage performance by employing the mode diversity. For instance, when $C_n^2=6\times10^{-15} \,\mathrm{m}^{-2/3}$ and $P_t=-4$ dBm, the outage probability is at a high level of $27\%$ for the channel with $i=\pm10$ in the absence of mode diversity. However, the corresponding outage probability in the presence of mode diversity is only $4\times 10^{-4}$. Similarly, When turbulence condition is $C_n^2=1\times10^{-15} \,\mathrm{m}^{-2/3}$ and seven modes are employed for multiplexing, the outage probability of the channel with $i=\pm5$ is $0.04$ for $P_t=-10$ dBm, however, the corresponding outage probability decreases to $1\times10^{-3}$ in the presence of mode diversity. 

Negative asymptotic slope of error probability or outage probability is usually used to characterize the diversity order of diversity systems \cite{majidcoop}. In SMM systems considered here since with the increase of $P_t$ the multiplexed channels turn to be interference-limited, error floors will occur for outage probability curves in high $P_t$ regime. As a result the conventional definition of diversity order is of no use. However, clear changes can be observed in the negative slopes of the performance curves at finite $P_t$ when mode diversity is employed. Therefore, one can still get some insights into the diversity gain using the normalized slopes of the outage probability curves with respect to that in the absence of mode diversity at finite $P_t$ \cite{majidcoop}.  For instance, in \figurename\textrm{ }\ref{SINR_pow7} one can calculate the normalized slopes as $1.74$, $4.98$ and $1.42$ for multiplexed channels with $i=\pm2$, $i=\pm5$ and $i=0$, respectively when $P_t=-10$ dBm. Hence, with this transmitted power the channels with $i=\pm5$ benefit the most from the mode diversity.
 
\subsection{$\epsilon$-Outage Achievable Rate}
Finally, let us investigate the achievable rates of the SMM system employing both multiplexing and mode diversity. The detected photon counts of the channel with transmitted mode state $i$ is given in (\ref{noitil}). After some algebraic manipulations, this expression can be rewritten as
\begin{equation}\label{outach}
\tilde{n}_{o,i}=\tilde{m}_{s,i}+\sqrt{\tilde{m}_{s,i}}\mathcal{Z}_s^{(i)}+\mathcal{Z}_0^{(i)},
\end{equation}
where 
\begin{align}
\tilde{m}_{s,i}&=\mu\rho_i^2\sum_{j\in\mathcal{M}_i}\beta_j^{(i)} |\alpha_{ij}|^2,\,\, \mathcal{Z}_0^{(i)}=\sum_{j\in\mathcal{M}_i}\beta_j^{(i)}Z_{0,j}^{(i)},\\\nonumber \mathcal{Z}_s^{(i)}&=\frac{1}{\sqrt{\sum_{j\in\mathcal{M}_i}\beta_j^{(i)} |\alpha_{ij}|^2}}\sum_{j\in\mathcal{M}_i}\beta_j^{(i)}|\alpha_{ij}|Z_{s,j}^{(i)}.
\end{align}
Since both $Z_{s,j}^{(i)}$ and $Z_{0,j}^{(i)}$ are zero mean Gaussian random variables, $\mathcal{Z}_s^{(i)}$ and $\mathcal{Z}_0^{(i)}$ are also zero mean Gaussian distributed with variance 
\begin{equation}
\sigma_{\mathcal{Z}_s^{(i)}}^2=\frac{\sum_{j\in\mathcal{M}}\left(\beta_j^{(i)}\right)^2|\alpha_{ij}|^2\sigma_{Z_{s,j}^{(i)}}^2}{\sum_{j\in\mathcal{M}}\beta_j^{(i)}|\alpha_{ij}|^2},\, \sigma_{\mathcal{Z}_0^{(i)}}^2=\sum_{j\in\mathcal{M}}\left(\beta_j^{(i)}\right)^2\sigma_{Z_{0,j}^{(i)}}^2,
\end{equation} 
where $\sigma_{Z_{s,j}^{(i)}}^2$ and $\sigma_{Z_{0,j}^{(i)}}^2$ are given in (\ref{var01}). The channel expression (\ref{outach}) is similar to that of the channel in the absence of mode diversity in (\ref{channelG})， where the average power constraint is now given by $E[\tilde{m}_{s,i}]=\mu P_t\sum_{j\in\mathcal{M}_i}\beta_j^{(i)} |\alpha_{ij}|^2/N$. Hence the achievable rate conditioned on the channel states can be expressed as 
	\begin{align}\label{rateDIV}
	\mathcal{C}_{i|\boldsymbol{\alpha_{i}}}=&\frac{1}{2}\mathrm{log}\frac{E[\tilde{m}_{s,i}]}{\sigma_{\mathcal{Z}_s^{(i)}}^2}+\frac{1}{2}\,\mathrm{log}\left(1+\frac{2\sigma_{\mathcal{Z}_s^{(i)}}^2}{E[\tilde{m}_{s,i}]}\right)-\frac{E[\tilde{m}_{s,i}]}{\sigma_{\mathcal{Z}_s^{(i)}}^2}-1\\\nonumber
	&+\frac{\sqrt{E[\tilde{m}_{s,i}]\left(E[\tilde{m}_{s,i}]+2\sigma_{\mathcal{Z}_s^{(i)}}^2\right)}}{\sigma_{\mathcal{Z}_s^{(i)}}^2}-\sqrt{\frac{\pi \sigma_{\mathcal{Z}_0^{(i)}}^2}{2E[\tilde{m}_{s,i}]\sigma_{\mathcal{Z}_s^{(i)}}^2}},
	\end{align}
			\begin{figure}[!t]
				\centering
				\includegraphics[width=0.55\textwidth]{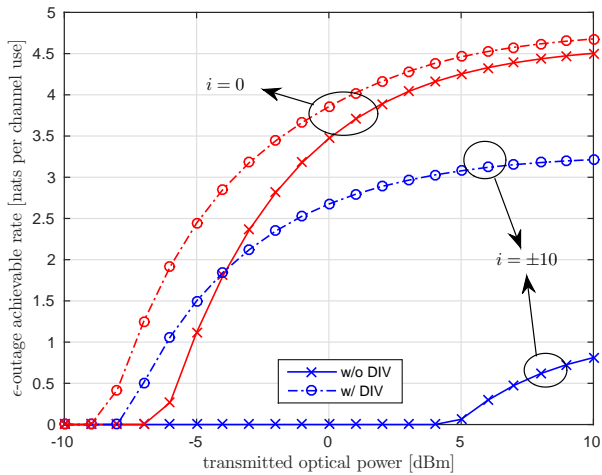}
				\caption{The $\epsilon$-outage achievable rate versus the transmitted optical power for OAM-based SMM system with and without the mode diversity where $\mathcal{N}=\{0,\pm10\}$, $C_n^2=6\times10^{-15} \,\mathrm{m}^{-2/3}$ and $\epsilon=0.01$.   }\label{ca_outage_pow3}
			\end{figure} 
where $\boldsymbol{\alpha_{i}}=[\alpha_{i1},\cdots,\alpha_{iM_i}]^T$ is the the vector of the instantaneous fadings of all combining branches. The $\epsilon$-outage achievable rate is defined as the largest rate $\mathcal{C}_\mathrm{out}$ that satisfies the condition \cite{Wilson}
	\begin{equation}\label{outC}
	\mathrm{Pr}\bigg\{\mathcal{C}_{i|\boldsymbol{\alpha_{i}}}<\mathcal{C}_\mathrm{out} \bigg\}<\epsilon,
	\end{equation}
where $\epsilon$ is a fixed value. $\epsilon$-outage achievable rate provides the maximum data rate that can be transmitted in the system under the condition that the outage criterion is satisfied. Using the optimal coefficients $\beta_j^{(i)}$ given in (\ref{beta}) which maximize the asymptotic output SINR of the SIMO link, the $1 \%$-outage achievable rate for a three-mode OAM-based SMM system is plotted in \figurename\textrm{ }\ref{ca_outage_pow3}. One can observe that using the mode diversity, the outage achievable rate can be significantly improved especially for the channel with mode state $i=\pm10$. For instance, when $P_t=5$ dBm, the outage achievable rate for $i=\pm10$ is negligible in the absence of mode diversity, however, when mode diversity is employed, more than $3$ nats per channel use outage achievable rate can be achieved. It is worth mentioning that although the use of  combining coefficients given by (\ref{beta}) can significantly improve the $\epsilon$-outage achievable rate, these coefficients do not maximize the receivable rate. Equation (\ref{beta}) is optimal in the sense of maximizing the asymptotic SINR (\ref{SNRasy}), however, the expression of the achievable rate given in (\ref{rateDIV}) is not a a direct function of SINR. Thus one might be able to find other coefficients which can achieve even higher $\epsilon$-outage achievable rates.       

\section{Conclusion}\label{conc}
In this paper, IM/DD SMM FSO systems with mutually coherent channels are investigated. Compared to the systems with mutually incoherent channels, the system considered here employs a single laser source with a narrow linewidth to generate different spatial modes, which simplifies the transmitter design and preserve the spectral DOFs. 
In order to evaluate the system performance justifiably, the average AAR is considered. For practical SMM systems, it is concluded that an optimal transmitted mode set with specific number of modes can be determined which maximizes the average asymptotic AAR. Moreover, under stronger turbulence, the number of modes in the optimal mode set decreases accordingly. In order to improve the reliability of every multiplexed channel in the system, we propose to use a mode diversity scheme which renders the SISO links in the system into SIMO links. The expression of the optimal combining coefficients is derived which maximizes the asymptotic SINR and the best diversity mode sets for different channels are discussed. Through outage performance analysis, it is concluded that using mode diversity, both the outage probability and $\epsilon$-outage achievable rate can be significantly improved. This technique is cost-effective and is a potential technique to improve the reliability of FSO SMM systems in the future.
    
	\bibliographystyle{IEEEtran}
	\bibliography{IEEEabrv,conference_paper}

\begin{thebibliography}{10}
\providecommand{\url}[1]{#1}
\csname url@samestyle\endcsname
\providecommand{\newblock}{\relax}
\providecommand{\bibinfo}[2]{#2}
\providecommand{\BIBentrySTDinterwordspacing}{\spaceskip=0pt\relax}
\providecommand{\BIBentryALTinterwordstretchfactor}{4}
\providecommand{\BIBentryALTinterwordspacing}{\spaceskip=\fontdimen2\font plus
\BIBentryALTinterwordstretchfactor\fontdimen3\font minus
  \fontdimen4\font\relax}
\providecommand{\BIBforeignlanguage}[2]{{%
\expandafter\ifx\csname l@#1\endcsname\relax
\typeout{** WARNING: IEEEtran.bst: No hyphenation pattern has been}%
\typeout{** loaded for the language `#1'. Using the pattern for}%
\typeout{** the default language instead.}%
\else
\language=\csname l@#1\endcsname
\fi
#2}}
\providecommand{\BIBdecl}{\relax}
\BIBdecl

\bibitem{willner}
A.~E. Willner, G.~Xie, L.~Li, Y.~Ren, Y.~Yan, N.~Ahmed, Z.~Zhao, Z.~Wang
  \emph{et~al.}, ``Design challenges and guidelines for free-space optical
  communication links using orbital-angular-momentum multiplexing of multiple
  beams,'' \emph{Journal of Optics}, vol.~18, no.~7, p. 074014, 2016.

\bibitem{wang}
J.~Wang, J.-Y. Yang, I.~M. Fazal, N.~Ahmed, Y.~Yan, H.~Huang, Y.~Ren, Y.~Yue,
  S.~Dolinar, M.~Tur \emph{et~al.}, ``Terabit free-space data transmission
  employing orbital angular momentum multiplexing,'' \emph{Nature Photonics},
  vol.~6, no.~7, pp. 488--496, 2012.

\bibitem{Anguita:08}
J.~A. Anguita, M.~A. Neifeld, and B.~V. Vasic, ``Turbulence-induced channel
  crosstalk in an orbital angular momentum-multiplexed free-space optical
  link,'' \emph{Appl. Opt.}, vol.~47, no.~13, pp. 2414--2429, May 2008.

\bibitem{zhao}
N.~Zhao, X.~Li, G.~Li, and J.~M. Kahn, ``Capacity limits of spatially
  multiplexed free-space communication,'' \emph{Nature Photonics}, vol.~9,
  no.~12, pp. 822--826, 2015.

\bibitem{Gibson:04}
G.~Gibson, J.~Courtial, M.~J. Padgett, M.~Vasnetsov, V.~Pas'ko, S.~M. Barnett,
  and S.~Franke-Arnold, ``Free-space information transfer using light beams
  carrying orbital angular momentum,'' \emph{Opt. Express}, vol.~12, no.~22,
  pp. 5448--5456, Nov 2004.

\bibitem{Paterson}
C.~Paterson, ``Atmospheric turbulence and orbital angular momentum of single
  photons for optical communication,'' \emph{Phys. Rev. Lett.}, vol.~94, p.
  153901, Apr 2005.

\bibitem{Gasulla:15}
I.~Gasulla and J.~M. Kahn, ``Performance of direct-detection
  mode-group-division multiplexing using fused fiber couplers,'' \emph{J.
  Lightwave Technol.}, vol.~33, no.~9, pp. 1748--1760, May 2015.

\bibitem{Huang:14}
H.~Huang, Y.~Cao, G.~Xie, Y.~Ren, Y.~Yan, C.~Bao, N.~Ahmed, M.~A. Neifeld,
  S.~J. Dolinar, and A.~E. Willner, ``Crosstalk mitigation in a free-space
  orbital angular momentum multiplexed communication link using $4\times4$
  {MIMO} equalization,'' \emph{Opt. Lett.}, vol.~39, no.~15, pp. 4360--4363,
  Aug 2014.

\bibitem{huang2015mode}
H.~Huang, G.~Milione, M.~P. Lavery, G.~Xie, Y.~Ren, Y.~Cao, N.~Ahmed, T.~A.
  Nguyen, D.~A. Nolan, M.-J. Li \emph{et~al.}, ``Mode division multiplexing
  using an orbital angular momentum mode sorter and {MIMO-DSP} over a
  graded-index few-mode optical fibre,'' \emph{Scientific reports}, vol.~5,
  2015.

\bibitem{survey}
M.~A. Khalighi and M.~Uysal, ``Survey on free space optical communication: A
  communication theory perspective,'' \emph{IEEE Communications Surveys
  Tutorials}, vol.~16, no.~4, pp. 2231--2258, Fourthquarter 2014.

\bibitem{Igarashi:15}
K.~Igarashi, D.~Souma, Y.~Wakayama, K.~Takeshima, Y.~Kawaguchi, T.~Tsuritani,
  I.~Morita, and M.~Suzuki, ``114 space-division-multiplexed transmission over
  9.8-km weakly-coupled-6-mode uncoupled-19-core fibers,'' in \emph{Optical
  Fiber Communication Conference Post Deadline Papers}.\hskip 1em plus 0.5em
  minus 0.4em\relax Optical Society of America, 2015, p. Th5C.4.

\bibitem{Ren:14}
Y.~Ren, G.~Xie, H.~Huang, N.~Ahmed, Y.~Yan, L.~Li, C.~Bao, M.~P.~J. Lavery
  \emph{et~al.}, ``Adaptive-optics-based simultaneous pre- and post-turbulence
  compensation of multiple orbital-angular-momentum beams in a bidirectional
  free-space optical link,'' \emph{Optica}, vol.~1, no.~6, pp. 376--382, Dec
  2014.

\bibitem{shengmei}
S.~M. Zhao, J.~Leach, L.~Y. Gong, J.~Ding, and B.~Y. Zheng, ``Aberration
  corrections for free-space optical communications in atmosphere turbulence
  using orbital angular momentum states,'' \emph{Opt. Express}, vol.~20, no.~1,
  pp. 452--461, Jan 2012.

\bibitem{Yadin:06}
Y.~Yadin and M.~Orenstein, ``Parallel optical interconnects over multimode
  waveguides,'' \emph{J. Lightwave Technol.}, vol.~24, no.~1, p. 380, Jan 2006.

\bibitem{Nazarathy:08}
M.~Nazarathy and A.~Agmon, ``Coherent transmission direct detection {MIMO} over
  short-range optical interconnects and passive optical networks,'' \emph{J.
  Lightwave Technol.}, vol.~26, no.~14, pp. 2037--2045, Jul 2008.

\bibitem{shane}
S.~M. Haas and J.~H. Shapiro, ``Capacity of wireless optical communications,''
  \emph{IEEE Journal on Selected Areas in Communications}, vol.~21, no.~8, pp.
  1346--1357, Oct 2003.

\bibitem{Yadin}
Y.~Yadin and M.~Orenstein, ``Parallel optical interconnects over multimode
  waveguides using mutually coherent channels and direct detection,''
  \emph{Journal of Lightwave Technology}, vol.~25, no.~10, pp. 3126--3131, Oct
  2007.

\bibitem{Legg}
P.~J. Legg, M.~Tur, and I.~Andonovic, ``Solution paths to limit interferometric
  noise induced performance degradation in ask/direct detection lightwave
  networks,'' \emph{Journal of Lightwave Technology}, vol.~14, no.~9, pp.
  1943--1954, Sep 1996.

\bibitem{Monroy}
I.~T. Monroy, E.~Tangdiongga, and H.~de~Waardt, ``On the distribution and
  performance implications of filtered interferometric crosstalk in optical
  {WDM} networks,'' \emph{Journal of Lightwave Technology}, vol.~17, no.~6, pp.
  989--997, Jun 1999.

\bibitem{Bikhazi}
N.~W. Bikhazi, M.~A. Jensen, and A.~L. Anderson, ``{MIMO} signaling over the
  mmf optical broadcast channel with square-law detection,'' \emph{IEEE
  Transactions on Communications}, vol.~57, no.~3, pp. 614--617, March 2009.

\bibitem{Arik:16}
S.~O. Arik and J.~M. Kahn, ``Direct-detection mode-division multiplexing in
  modal basis using phase retrieval,'' \emph{Opt. Lett.}, vol.~41, no.~18, pp.
  4265--4268, Sep 2016.

\bibitem{ArikarXiv}
S.~O. {Arik} and J.~M. {Kahn}, ``{Low-complexity implementation of convex
  optimization-based phase retrieval},'' \emph{ArXiv e-prints}, Jul. 2017.

\bibitem{majidNF}
M.~Safari and S.~Hranilovic, ``Diversity and multiplexing for near-field
  atmospheric optical communication,'' \emph{IEEE Transactions on
  Communications}, vol.~61, no.~5, pp. 1988--1997, May 2013.

\bibitem{Anguita_pdf}
J.~A. Anguita, M.~A. Neifeld, and B.~V. Vasic, ``Modeling channel interference
  in an orbital angular momentum-multiplexed laser link,'' pp. 74\,640U1--6,
  2009.

\bibitem{lee_con}
E.~J. Lee and V.~W.~S. Chan, ``Diversity coherent receivers for optical
  communication over the clear turbulent atmosphere,'' in \emph{2007 IEEE
  International Conference on Communications}, June 2007, pp. 2485--2492.

\bibitem{pream}
M.~Razavi and J.~H. Shapiro, ``Wireless optical communications via diversity
  reception and optical preamplification,'' \emph{IEEE Transactions on Wireless
  Communications}, vol.~4, no.~3, pp. 975--983, May 2005.

\bibitem{Shah}
A.~Shah and A.~M. Haimovich, ``Performance analysis of maximal ratio combining
  and comparison with optimum combining for mobile radio communications with
  cochannel interference,'' \emph{IEEE Transactions on Vehicular Technology},
  vol.~49, no.~4, pp. 1454--1463, Jul 2000.

\bibitem{gagliardi}
R.~M. Gagliardi and S.~Karp, ``Optical communications,'' \emph{New York,
  Wiley-Interscience, 1976. 445 p.}, vol.~1, 1976.

\bibitem{Humblet}
P.~A. Humblet and M.~Azizoglu, ``On the bit error rate of lightwave systems
  with optical amplifiers,'' \emph{Journal of Lightwave Technology}, vol.~9,
  no.~11, pp. 1576--1582, Nov 1991.

\bibitem{Li}
T.~Li and M.~C. Teich, ``Photon point process for traveling-wave laser
  amplifiers,'' \emph{IEEE Journal of Quantum Electronics}, vol.~29, no.~9, pp.
  2568--2578, Sep 1993.

\bibitem{Moser}
S.~M. Moser, ``Capacity results of an optical intensity channel with
  input-dependent gaussian noise,'' \emph{IEEE Transactions on Information
  Theory}, vol.~58, no.~1, pp. 207--223, Jan 2012.

\bibitem{numerical}
J.~D. Schmidt, ``Numerical simulation of optical wave propagation with examples
  in {MATLAB}.''\hskip 1em plus 0.5em minus 0.4em\relax SPIE Bellingham, WA,
  2010.

\bibitem{Anguita:co}
J.~A. Anguita, J.~Herreros, and I.~B. Djordjevic, ``Coherent multimode oam
  superpositions for multidimensional modulation,'' \emph{IEEE Photonics
  Journal}, vol.~6, no.~2, pp. 1--11, April 2014.

\bibitem{Ren:16}
Y.~Ren, Z.~Wang, G.~Xie, L.~Li, A.~J. Willner, Y.~Cao, Z.~Zhao, Y.~Yan
  \emph{et~al.}, ``Atmospheric turbulence mitigation in an {OAM}-based {MIMO}
  free-space optical link using spatial diversity combined with {MIMO}
  equalization,'' \emph{Opt. Lett.}, vol.~41, no.~11, pp. 2406--2409, Jun 2016.

\bibitem{shenjie}
S.~Huang and M.~Safari, ``Spatial-mode multiplexing with zero-forcing
  beamforming in free space optical communications,'' in \emph{2017 IEEE
  International Conference on Communications Workshops (ICC Workshops)}, May
  2017, pp. 331--336.

\bibitem{Mehrpoor}
G.~R. Mehrpoor, M.~Safari, and B.~Schmauss, ``Free space optical communication
  with spatial diversity based on orbital angular momentum of light,'' in
  \emph{2015 4th International Workshop on Optical Wireless Communications
  (IWOW)}, Sept 2015, pp. 78--82.

\bibitem{Navidpour}
S.~M. Navidpour, M.~Uysal, and M.~Kavehrad, ``{BER} performance of free-space
  optical transmission with spatial diversity,'' \emph{IEEE Transactions on
  Wireless Communications}, vol.~6, no.~8, pp. 2813--2819, August 2007.

\bibitem{shin}
H.~Shin and M.~Z. Win, ``{MIMO} diversity in the presence of double
  scattering,'' \emph{IEEE Transactions on Information Theory}, vol.~54, no.~7,
  pp. 2976--2996, July 2008.

\bibitem{Hasna}
M.~O. Hasna and M.~S. Alouini, ``Harmonic mean and end-to-end performance of
  transmission systems with relays,'' \emph{IEEE Transactions on
  Communications}, vol.~52, no.~1, pp. 130--135, Jan 2004.

\bibitem{Charash}
U.~Charash, ``Reception through {Nakagami} fading multipath channels with
  random delays,'' \emph{IEEE Transactions on Communications}, vol.~27, no.~4,
  pp. 657--670, Apr 1979.

\bibitem{relay}
M.~Safari and M.~Uysal, ``Relay-assisted free-space optical communication,''
  \emph{IEEE Transactions on Wireless Communications}, vol.~7, no.~12, pp.
  5441--5449, December 2008.

\bibitem{majidcoop}
------, ``Cooperative diversity over log-normal fading channels: performance
  analysis and optimization,'' \emph{IEEE Transactions on Wireless
  Communications}, vol.~7, no.~5, pp. 1963--1972, May 2008.

\bibitem{Wilson}
S.~G. Wilson, M.~Brandt-Pearce, Q.~Cao, and J.~H. Leveque, ``Free-space optical
  {MIMO} transmission with {Q}-ary {PPM},'' \emph{IEEE Transactions on
  Communications}, vol.~53, no.~8, pp. 1402--1412, Aug 2005.

\end{thebibliography}
	
\end{document}